\documentclass[twocolumn,amsmath,amssymb,prd]{revtex4}

\def\ga{\gamma}

\def\Ga{\Gamma}
\def\De{\Delta}

\def\La{\Lambda}

\def\cA{{\cal A}}

\def\fr#1#2{{{#1} \over {#2}}}
\def\half{{\textstyle{1\over 2}}}

\def\lsim{\mathrel{\rlap{\lower4pt\hbox{\hskip1pt$\sim$}}
    \raise1pt\hbox{$<$}}}
\def\gsim{\mathrel{\rlap{\lower4pt\hbox{\hskip1pt$\sim$}}
    \raise1pt\hbox{$>$}}}

\def\etal{{\it et al.}}

\newcommand{\beq}{\begin{equation}}
\newcommand{\eeq}{\end{equation}}
\newcommand{\bea}{\begin{eqnarray}}
\newcommand{\eea}{\end{eqnarray}}
\newcommand{\bit}{\begin{itemize}}
\newcommand{\eit}{\end{itemize}}
\newcommand{\rf}[1]{(\ref{#1})}

\def\Re{\hbox{Re}\,}
\def\Im{\hbox{Im}\,}

\def\bb{\overline{B}{}}

\def\kkb{$K^0$-$\overline{K}{}^0$}
\def\ddb{$D^0$-$\overline{D}{}^0$}
\def\bdbdb{$B_d^0$-$\overline{B}{}_d^0$}
\def\bsbsb{$B_s^0$-$\overline{B}{}_s^0$}

\def\dmd{\De m_d}
\def\dgd{\De \Ga_d}
\def\dms{\De m_s}
\def\dgs{\De \Ga_s}
\def\dmD{\De m_D}
\def\dgD{\De \Ga_D}

\def\da{\De a{}}
\def\dak{(\De a^{K}){}}
\def\dad{(\De a^{D}){}}
\def\dabd{(\De a^{B_d}){}}
\def\dabs{(\De a^{B_s}){}}

\begin{document}

\title{CPT violation and $\boldmath B$-meson oscillations}

\author{V.\ Alan Kosteleck\'y and Richard J.\ Van Kooten}

\affiliation{Physics Department, Indiana University,
Bloomington, IN 47405, U.S.A.}

\date{IUHET 545, July 2010; 
{\it Rapid Communications, Physical Review D}, in press}

\begin{abstract}

Recent evidence for anomalous CP violation 
in $B$-meson oscillations 
can be interpreted as resulting from CPT violation.
This yields the first sensitivity 
to CPT violation in the $B_s^0$ system,
with the relevant coefficient for CPT violation 
constrained at the level of parts in $10^{12}$.

\end{abstract}

\maketitle

Experimental studies of spacetime symmetries
involving the discrete transformations
under charge conjugation C,
parity inversion P,
time reversal T,
and their products CP and CPT have played a major role
in establishing the Standard Model (SM) of particle physics.
All these symmetries are known to be broken except CPT, 
and their description in terms of the SM
has been in excellent agreement with laboratory experiments.
Among the most powerful tools
available for investigations of these symmetries
are the neutral-meson systems,
in which particles and antiparticles mix
interferometrically and thereby offer
high sensitivity to deviations from exact symmetry.

The D0 Collaboration
has recently presented data supporting 
an anomalous like-sign dimuon charge asymmetry
in $B$-meson mixing
\cite{d01,d02},
interpreting it as evidence for 
CPT-invariant CP violation beyond the SM.
Here,
we show that this anomalous asymmetry
could also arise from T-invariant CP violation
in \bsbsb\ mixing.
A CPT-violating effect in $B$-meson mixing
was predicted some time ago
\cite{kp}
as potentially arising from spontaneous breaking of Lorentz symmetry
in an underlying unified theory 
\cite{ksp},
and the usual requirement of CPT-invariant CP violation 
for baryogenesis
\cite{ads}
can be evaded in this context
\cite{bsyn}.
The \bsbsb\ system is of particular interest
for studies of CPT violation
because several complete particle-antiparticle oscillations
occur within a meson lifetime
\cite{d0cdf}.
As part of the analysis here,
we show that the anomalous like-sign dimuon charge asymmetry
offers sensitivity to CPT breaking,
and we use this asymmetry to obtain the first quantitative measure 
of CPT violation in the \bsbsb\ system.

An appropriate framework for investigating
CPT violation in neutral mesons
is effective field theory.
In this context,
CPT violation is necessarily accompanied by Lorentz violation
\cite{owg}.
We can therefore work here within
the comprehensive effective field theory
describing general Lorentz violation at attainable energies
known as the Standard-Model Extension (SME)
\cite{ck}.
Each CPT-violating term in the SME Lagrange density 
is the product of a CPT-violating operator
and a controlling coefficient.
The SME contains both the SM and General Relativity,
so it serves as a realistic theory 
for analyzing experimental data for signals of CPT violation.
Several SME-based searches for CPT violation 
with neutral-meson oscillations 
\cite{kloe,ktev,focus,babar}
and numerous investigations 
using a wide variety of other physical systems
\cite{tables}
have been performed over the past decade.

The analysis of meson mixing in the SME context 
reveals the four neutral-meson systems
\kkb, \ddb, \bdbdb, and \bsbsb\
contain a total of 16 independent observables for CPT violation
\cite{ak3}.
The corresponding 16 combinations of SME coefficients 
are conventionally denoted as
$\dak_\mu$,
$\dad_\mu$,
$\dabd_\mu$,
$\dabs_\mu$.
These coefficients are known to be observable
only in flavor-changing experiments with neutral mesons or neutrinos
\cite{neutrinos}
and in gravitational experiments
\cite{akjt}.
Several experimental searches 
have yielded high sensitivities to 
certain components of
$\dak_\mu$, $\dad_\mu$, and $\dabd_\mu$
\cite{kloe,ktev,focus,babar}.
In this work,
we report the first sensitivity to the coefficient $\dabs_\mu$.
We also outline a procedure that could improve on this result
using the full D0 dataset. 

The D0 Collaboration measures the dimuon charge asymmetry
\beq
A^b_{\rm sl} = 
\fr{N_b^{++} - N_b^{--}}{N_b^{++} + N_b^{--}} ,
\label{dimuonchargeasymm}
\eeq
where $N_b^{++}$ and $N_b^{--}$
represent the number of events
in which two $b$ hadrons decay semileptonically
into two positive muons and two negative muons,
respectively.
One measurement of this asymmetry
is obtained by correcting the raw like-sign dimuon sample
for various backgrounds,
yielding
\cite{d02} 
\beq
A^b_{\rm sl} = 
-0.00736 \pm 0.00266 \pm 0.00305,
\label{dimuonasymm}
\eeq
where the first error is statistical and the second systematic.
Combining in quadrature yields 
an effect at 1.8 standard deviations.
The D0 Collaboration also studies 
the inclusive `wrong-charge' muon charge asymmetry 
$a^b_{\rm sl}$
of semileptonic decays of $b$ hadrons
to muons with charge opposite to that of the original $b$ quark,
\beq
a^b_{\rm sl} =
\fr{ \Ga(\bb \to \mu^+ X) - \Ga(B \to \mu^- X) }
{ \Ga(\bb \to \mu^+ X) + \Ga(B \to \mu^- X) }.
\label{asymmT}
\eeq
This asymmetry is a measure of CPT-invariant CP violation
and hence of T violation.
Assuming CPT symmetry holds
and under other mild assumptions 
such as no direct CP violation,
it can be shown that $A^b_{\rm sl} = a^b_{\rm sl}$ 
\cite{gnr},
which enables a second measurement of $A^b_{\rm sl}$.
This second measurement is consistent 
with no effect at 0.4 standard deviations. 
The final D0 result for $A^b_{\rm sl}$
is obtained by combining the two measurements 
to minimize systematic uncertainties.
It reveals a signal 3.2 standard deviations
away from the SM prediction for CPT-preserving T violation,
which is \cite{ln,d02}
\beq
A^b_{\rm sl} {\rm (SM)} =
(-2.3^{+0.5}_{-0.6}) \times 10^{-4}.
\label{TasymmSM}
\eeq
For our present purposes,
the second measurement using the asymmetry \rf{asymmT}
and the combined measurement both turn out to be irrelevant,
so only the first result \rf{dimuonasymm} for $A_{\rm sl}^b$
is involved in the analysis that follows.

In this work,
we allow for T-invariant CP violation
in \bsbsb\ oscillations. 
A measure of this CPT violation
is given by the inclusive `right-charge' muon 
charge asymmetry $\cA^b_{\rm CPT}$
of semileptonic decays of $b$ hadrons
to muons with the same charge 
as that of the original $b$ quark,
\beq
\cA^b_{\rm CPT} =
\fr{ \Ga(\bb \to \mu^- X) - \Ga(B \to \mu^+ X) }
{ \Ga(\bb \to \mu^- X) + \Ga(B \to \mu^+ X) }.
\label{asymmCPT}
\eeq
In terms of this CPT asymmetry and the T asymmetry \rf{asymmT},
we find the dimuon charge asymmetry $A^b_{\rm sl}$
of Eq.\ \rf{dimuonchargeasymm} can be written 
in the nested form 
\bea
A^b_{\rm sl} &=&
{
\displaystyle
\left( \fr{1 + a^b_{\rm sl}}{1 - a^b_{\rm sl}}
-
\fr{1 + \cA^b_{\rm CPT}}{1 - \cA^b_{\rm CPT}}
\right)
\over
\displaystyle
\left(
\fr{1 + a^b_{\rm sl}}{1 - a^b_{\rm sl}}
+
\fr{1 + \cA^b_{\rm CPT}}{1 - \cA^b_{\rm CPT}}
\right)
}
\approx
a^b_{\rm sl} - \cA^b_{\rm CPT},
\label{nested}
\eea
where the last expression
assumes small T and CPT violation
at first order in the asymmetries.
This expression reveals 
that the dimuon charge asymmetry $A^b_{\rm sl}$
is sensitive to CPT violation
as well as T violation.
In what follows,
the result \rf{nested} is used 
to obtain the first quantitative measure
of CPT violation in the \bsbsb\ system. 

For definiteness,
we assume the only source of T violation
is the SM contribution
$a^b_{\rm sl} {\rm (SM)} = A^b_{\rm sl} {\rm (SM)}$
given by Eq.\ \rf{TasymmSM}.
Combining with the D0 dimuon asymmetry \rf{dimuonasymm}
yields the value 
\beq
\cA^b_{\rm CPT} = 
0.00713 \pm 0.00405,
\label{asymmexpt}
\eeq
where the D0 statistical and systematic errors
are combined in quadrature.
Our goal is to interpret this result as a measure
of CPT violation in $B$-meson mixing,
and in particular in the \bsbsb\ system.

In general,
oscillations of neutral mesons
are governed by a $2\times 2$ effective hamiltonian $\La$
\cite{kdgh}.
The CPT-violating contributions to $\La$
are controlled by the difference $\De\La =\La_{11} - \La_{22}$
of the diagonal terms of $\La$,
while the off-diagonal terms govern T violation.
The size of CPT violation 
is unknown \it a priori. \rm 
We adopt here the $w\xi$ formalism for $\La$
\cite{ak3},
which is independent of phase conventions
and allows for CPT violation of arbitrary size.
In this formalism,
CPT violation is governed by a complex parameter $\xi$
of any magnitude,
and
$\De\La = - (\De m + \half i \De \Ga) \xi$.
For the \bsbsb\ system,
$\De m \equiv \dms = m_H - m_L$
is the mass difference between the heavy and light eigenstates,
$\De \Ga \equiv \dgs = \Ga_L - \Ga_H$
is their width difference,
and the parameter for CPT violation is denoted $\xi_s$.
We also adopt the standard notation
$x_s = \dms/\Ga_s$,
$y_s = \dgs/2\Ga_s$,
$2\Ga_s = \Ga_L + \Ga_H$.

Since CPT violation comes with Lorentz violation
\cite{owg},
the complex parameter $\xi$ cannot be a scalar.
Instead,
it must depend on the meson 4-momentum
and is therefore a frame-dependent quantity.
For example,
the rotation of the Earth 
relative to the constant vector $\De\vec a$
typically generates a variation with sidereal time in $\xi$ 
\cite{ak1}.
The canonical frame used in studies of CPT and Lorentz violation
is the Sun-centered frame with coordinates $(T,X,Y,Z)$
\cite{km}.
In this frame,
the CPT-violating parameter 
$\xi \equiv \xi(T, \vec p, \De a_\mu)$
is a function of sidereal time $T$,
meson 4-momentum $(E(\vec p),\vec p)$,
and the four constant SME coefficients $\De a_\mu$ 
for CPT violation for the given meson system.
The explicit functional form of $\xi$ 
can be found using perturbation theory for the SME
and is given as Eq.\ (14) of Ref.\ \cite{ak3}.
Hermiticity of the Lagrange density
ensures the reality of $\De\La$,
which for the \bsbsb\ system implies the condition
$y_s\Re\xi_s + x_s\Im\xi_s = 0$.

For our present purposes,
it suffices to average over the sidereal time
and the meson 4-momentum spectrum.
Since the particle distributions 
from $b$-hadron decay for the Fermilab collider 
are symmetric in local detector polar coordinates for D0,
the dependence on spatial components $\dabs_J$
cancels through this procedure.
We obtain the averaged value
\beq
\overline {\Im \xi_s} = 
\fr {y_s}{x_s^2 +y_s^2} 
\fr {{\overline{\ga}} \dabs_T} {\Ga_s} ,
\label{xiav}
\eeq
where $\overline{\ga}\simeq 4.1$ 
is the mean gamma boost factor for the $B_s^0$ mesons
in the D0 experiment.

Given the result \rf{xiav}, 
we can extract a measurement of $\dabs_\mu$
from the value \rf{asymmexpt}
once an expression for $\cA^b_{\rm CPT}$ is known 
in terms of $\overline{\Im \xi_s}$.
To derive this relationship for $\cA^b_{\rm CPT}$,
we note that  
\beq
\cA^b_{\rm CPT} = 
\fr {R^- - R^+} {R^- + R^+},
\label{asymmCPTnum}
\eeq
where $R^\pm$ represents the number of right-sign decays
into $\mu^\pm X$.
As measured at D0,
these quantities are a sum
over contributions from the \bdbdb\ system,
from the \bsbsb\ system,
and from all other $b$ hadrons.
Labeling these three sources as $q = d$, $s$, $u$
and using an overbar to identify
quantities for the $b$ quark,
we can write
\cite{gnr}
\bea
R^+ &=& 
f_d T_d \Ga^{\rm sl}_d
+ f_s T_s \Ga^{\rm sl}_s
+ f_u T_u \Ga^{\rm sl}_u,
\nonumber\\
R^- &=& 
\overline f_d \overline T_d \overline \Ga{}^{\rm sl}_d
+\overline f_s \overline T_s \overline \Ga{}^{\rm sl}_s
+\overline f_u \overline T_u \overline \Ga{}^{\rm sl}_u,
\eea
where we denote the production fractions as $f_q$,
the time-integrated probabilities 
for $B\to B$, $\overline{B}\to\overline{B}$,
or direct decay of nonmixing states as $T_q$,
and the semileptonic decay rates as $\Ga^{\rm sl}_q$.

Taking for definiteness 
zero direct T and CPT violation in semileptonic decays,
we have $\overline \Ga{}^{\rm sl}_q = \Ga^{\rm sl}_q$.
It is also a reasonable approximation to take
$\Ga^{\rm sl}_d = \Ga^{\rm sl}_s = \Ga^{\rm sl}_u$.
Symmetric production implies $\overline f_q = f_q$,
while $f_u = 1 - f_d - f_s$.
The absence of mixing for $q=u$ implies 
$T_u = 1/\Ga_u$,
where $\Ga_u$ is the total decay rate 
for the nonmixing $b$ hadrons,
which include the $B^\pm$ mesons and the $b$ baryons.
The time-dependent mixing and decay probabilities 
for neutral $B$ mesons in the $w\xi$ formalism
are given explicitly as Eq.\ (19) of Ref.\ \cite{ak3}.
These can be integrated over all time $t$
to yield $T_d$ and $T_s$.

For simplicity,
suppose the only source of CPT violation
comes from \bsbsb\ mixing.
Then,
integrating the probability for $B_d^0 \to B_d^0$ 
over all time $t$ gives $T_d = {z_{d+}}/ {2\Ga_d}$,
while the integration for $B_s^0 \to B_s^0$ yields 
\beq
T_s = \fr 1 {2\Ga_s} 
(z_{s+} 
+ 2 z_{s-} x_s \overline{\Im \xi_s}
+ z_{s-} z_{s0} (\overline{\Im \xi_s})^2).
\eeq
In these equations,
we define
\beq
z_{q\pm} = \fr 1 {(1-y_q^2)} \pm \fr 1 {(1+x_q^2)},
\quad
z_{q0} = (x_q^2 + y_q^2)/y_q^2.
\eeq
Applying CPT gives the additional relations $\overline T_d = T_d$
and $\overline T_s = T_s(\xi_s \to -\xi_s)$.

Collecting the results,
we finally obtain the CPT asymmetry 
\beq
\cA^b_{\rm CPT} =
\fr{2 f_s z_{s-} x_s \overline{\Im \xi_s} }
{ f_d z_{d+} + f_s z_{s+} + 2 f_u
+ f_s z_{s-} z_{s0} (\overline{\Im \xi_s})^2}.
\label{asymmtheory}
\eeq
We remark in passing
that the form of this result for $\cA^b_{\rm CPT}$
holds also in the unaveraged case,
provided Eq.\ \rf{xiav} is replaced
with the complete expression for 
$\Im\xi_s(T, \vec p, \dabs_\mu)$
and the reasonable approximation is made
that the decays occur over times $t$ negligible
compared to the sidereal variation with $T$.

To match the theoretical expression \rf{asymmtheory}
to the result \rf{asymmexpt} obtained from the D0 experiment,
we adopt the values
$x_d = 0.774 \pm 0.008$,
$y_d = 0$,
$x_s = 26.2 \pm 0.5$,
$y_s = 0.046 \pm 0.027$,
$f_d = 0.323 \pm 0.037$,
and $f_s = 0.118 \pm 0.015$
\cite{d02,pdg}.
Inverting the expression \rf{asymmtheory}
yields
\beq
\overline {\Im \xi_s} = (2.3 \pm 1.7)\times 10^{-3}.
\label{imxiresult}
\eeq
We can also extract the desired measurement 
of the SME coefficient $\dabs_T$ for CPT violation,
which is
\beq
\dabs_T = 
(3.7 \pm 3.8) \times 10^{-12} {\rm ~GeV},
\label{result}
\eeq
where $\Ga_s = (4.47 \pm 0.08)\times 10^{-13}$ GeV
\cite{pdg}.
This corresponds to the bound 
\beq
- 3.8 \times 10^{-12} < \dabs_T < 1.1 \times 10^{-11} 
\label{confint}
\eeq
at the 95\% confidence level.

The value \rf{result}
represents the first sensitivity to 
CPT violation in the \bsbsb\ system.
The result \rf{imxiresult} for $\Im \xi_s$
is consistent with no effect at 1.4 standard deviations,
which is a reasonable result 
given the size of the systematic errors
in the basic D0 asymmetry 
\rf{dimuonasymm}
and the SM-corrected asymmetry 
\rf{asymmexpt}.
The fractional error on the coefficient $\dabs_T$ 
for CPT violation is greater,
due primarily to the comparatively large uncertainty 
in the value of $y_s$.

For the D0 study of CPT-invariant CP violation,
the signal of 3.2 standard deviations
was obtained by reducing the systematics on 
$A^b_{\rm sl}$ by combining the result \rf{dimuonasymm}
with an independent measurement of $a^s_{\rm sl}$.
We observe here that
a similar technique could be used
in the present context of CPT violation.
The basic idea is to reduce the systematics 
by combining the result \rf{dimuonasymm} for $A^b_{\rm sl}$ 
with an independent measurement of CPT violation.
The relevant quantity for the latter measurement 
is the asymmetry $\cA^b_{\rm CPT}$
for inclusive `right-charge' muon semileptonic decays
defined in Eq.\ \rf{asymmCPT}.
The overall CPT reach including the result \rf{result}
might be substantially sharpened via this method.
However,
extracting the asymmetry $\cA^b_{\rm CPT}$
requires access to the full D0 dataset
and hence lies outside our scope. 

For a neutral meson containing valence quark $q_1$
and antiquark $q_2$,
the observable $\da_\mu$ is given by
$\da_\mu \approx r_{q_1}a^{q_1}_\mu - r_{q_2}a^{q_2}_\mu$.
The coefficients $a^{q_1}_\mu$, $a^{q_2}_\mu$
appear in the SME Lagrange density
in terms of the form $- a^q_\mu \overline{q} \ga^\mu q$
for each quark $q$,
while $r_{q_1}$ and $r_{q_2}$
are quantities of order one
arising from quark-binding and normalization effects 
\cite{kp}.
The value of $\De a_\mu$ 
is then primarily determined by the heaviest valence quark.
Note this implies that the zero-sum rule
\beq
\dak_\mu - \dabd_\mu + \dabs_\mu \approx 0
\label{zerosum}
\eeq
holds to a good approximation.

For \bdbdb\ mixing,
the BaBar Collaboration has obtained the measurement
\cite{babar}
\bea
\dabd_T - 0.30 \dabd_Z &=& 
\nonumber\\
&&
\hskip -100pt
(-3.0 \pm 2.4) \times 10^{-15} (\dmd/\dgd) {\rm ~GeV}.
\label{babarresult}
\eea
The ratio $\dmd/\dgd \gsim 10.6$ in this case
\cite{pdg},
so this measurement is compatible 
with the result \rf{result}
and the zero-sum rule \rf{zerosum}.
For the \ddb\ system,
the FOCUS Collaboration has obtained the measurement
\cite{focus}
\bea
\dad_T - 0.60 \dad_Z &=& 
\nonumber\\
&&
\hskip -80pt
(1.8 \pm 3.0) \times 10^{-16} (\dmD/\dgD) {\rm ~GeV}.
\label{focusresult}
\eea
The ratio $\dmD/\dgD \simeq 0.6$ is smaller here,
yielding an improved sensitivity 
\cite{pdg},
albeit to effects involving other quark flavors
and SME coefficients.
Several results have also been obtained for the \kkb\ system.
By studying different processes,
the KLOE Collaboration has obtained the independent measurements
\cite{kloe}
\bea
\dak_T &=& 
(0.4 \pm 1.8) \times 10^{-17} {\rm ~GeV},
\nonumber\\
\dak_Z &=& 
(2.4 \pm 9.7) \times 10^{-18} {\rm ~GeV}.
\label{kloeresult}
\eea
Using data from the Fermilab E773 experiment
\cite{e773},
a constraint of 
\beq
| \dak_T - 0.60 \dak_Z | 
\lsim 5\times 10^{-21} {\rm ~GeV}
\label{e773result}
\eeq
has also been obtained
\cite{ak1}.
These values are all compatible 
with the result \rf{result}
and the zero-sum rule \rf{zerosum}.

More general analyses of the D0 data could 
in principle be countenanced.
Given sufficient statistics
and a good understanding of the spectrum,
the spatial coefficients $\dabs_J$ for CPT violation
could be measured and disentangled 
by combining a search for sidereal variations
with spectral analysis.
Sidereal sensitivities have already
been obtained by 
KTeV \cite{ktev},
KLOE \cite{kloe},
FOCUS \cite{focus},
and 
BaBar \cite{babar}.
All the sidereal results 
are compatible with the result \rf{result}.

Another option for future investigation
is to allow for nonzero contributions
from $\Im \xi_d$ in the \bdbdb\ system
simultaneously with ones from $\Im \xi_s$.
This requires a nonzero value of $y_d$.
With both effects present,
and averaging over 4-momentum and sidereal time as before,
the asymmetry \rf{asymmtheory} acquires a term
in the numerator proportional to $\overline{\Im \xi_d}$,
while the denominator contains an additional term
proportional to $(\overline{\Im \xi_d})^2$.
This analysis would therefore yield a constraint involving
both $\dabd_T$ and $\dabs_T$,
albeit with a large error 
due to the current uncertainty in the value of $y_d$. 
A global fit of this type 
could also combine data from different experiments 
for the \bdbdb\ and \bsbsb\ systems.
Ideally,
information from \kkb\ experiments 
would be incorporated via Eq.\ \rf{zerosum} 
to extend further the CPT reach. 

Analyses along these lines are also well suited
to other ongoing experiments investigating neutral mesons.
Searches with high statistics and high boost,
such as those feasible at the LHCb experiment 
\cite{lhcb}
with average boost factor $\overline\ga \simeq 15$-$20$,
offer the capability to study CPT violation in $B$ mesons
with sensitivities unattained to date. 
The results presented here
outline a potential window 
for the exploration of physics 
beyond the SM
and can serve as an impetus for future studies
of CPT violation.

\vskip 0.1 truein

This work was supported in part
by the Department of Energy
under grant DE-FG02-91ER40661
and by the Indiana University Center for Spacetime Symmetries.

\end{document}